\documentclass[a4paper,11pt]{article}
\usepackage{jinstpub} 
\usepackage{lineno}
\usepackage{subcaption}
\usepackage[separate-uncertainty=true]{siunitx}
\DeclareSIUnit{\neqpcm}{\si{n_{eq}\per\centi\meter\squared}}

\title{Fast Timing and TCT Measurement Results of CNM Double-Sided 3D Sensors }

\author[a,1]{F. Lex,\note{Corresponding author.}}
\author[b]{A. Dierlamm,}
\author[a]{T. Hasler,}
\author[b]{L. Hahn,}
\author[a]{M. Hauser,}
\author[a]{K. Jakobs,}
\author[a]{R. Koppenhöfer,}
\author[a]{O. Kreyscher,}
\author[a]{U. Parzefall,}
\author[b]{N. Radhakrishnan,}
\author[b]{B. Regnery,}
\author[a]{Y. Sibold,}
\author[a]{D. Sperlich,}
\author[b]{L. Stockmeier}
\author[a]{and I. Zatocilova}
\affiliation[a]{Albert-Ludwigs-Universität Freiburg,\\Hermann-Herder-Str. 3, 79104 Freiburg i. Br., Germany}
\affiliation[b]{Karlsruhe Institute of Technology,\\Kaiserstraße 12, 76131 Karlsruhe, Germany}

\emailAdd{fabian.simon.lex@cern.ch}

\abstract{Sensors with fast timing capabilities (on the order of tens of picoseconds) are a critical component of future HEP tracking detectors, providing the ability to disentangle high multiplicity events. Silicon 3D sensors are capable of delivering this temporal resolution and additionally display an excellent radiation hardness, suitable for the harsh environments expected in future hadron colliders (e.g. FCC-hh). 	
In the course of the RD50 project "3D detectors optimized for timing applications", double-sided 3D sensors with two different column layouts, hexagonal and orthogonal, and different column counts were designed and produced by CNM.

In this study, results of IV, top-TCT and source timing measurements of unirradiated as well as proton irradiated sensor structures are presented. The IV characterisation shows a strong decrease of the overall high leakage current of the sensors after irradiation. Top-TCT measurements with both an infrared and red laser reveal a mostly homogenous charge collection in the sensor bulk, while higher signals are observed in the outermost columns for surface charge injection. The time resolution of sensors irradiated up to a \SI{1}{\mega \electronvolt} neutron equivalent fluence of \SI{3e15}{\per \centi \meter \squared} reaches down to ca. \SI{65}{\pico \second}, showing only a slight degradation compared to unirradiated sensors.}


\begin{document}
\maketitle
\flushbottom

\section{Introduction}
\label{sec:intro}

With the high energy physics (HEP) community shifting their focus from discovery to precision, ever more stringent requirements are placed on the detectors. For silicon tracking detectors, which are a critical part of all current and future HEP experiments, this is reflected in demands for higher radiation tolerance, rate capability, spatial and, in order to disentangle tracks in high pileup collisions, time resolution. The required time resolution of a few tens of picoseconds cannot be achieved with regular planar sensors due to their large electrode separation. The currently prevalent technologies for fast timing applications are Low-Gain-Avalanche Diodes (LGADs)~\cite{LGAD}, offsetting the smaller collected charge from thinner sensors with internal gain, and 3D sensors~\cite{Parker_2011}, reducing the electrode distance by utilizing n$^+$ and p$^+$ doped columns in the detector bulk. The latter allows for the same sensor thickness, and therefore collected signal, as in regular planar sensors, while having a smaller depletion voltage and shorter drift distances for any created charges, leading to improved radiation hardness and timing characteristics. In this study, IV, top transient current technique (top-TCT) and $\beta$-source timing results of unirradiated as well as irradiated double-sided 3D sensors produced by the Centro Nacional de Microelectrónica (CNM) are presented.

\section{Sensors \& Setups}
\subsection{CNM Double-Sided 3D Sensors}
\label{sec:sensors}

\begin{figure}[htbp]
	\begin{subfigure}[f]{0.4\textwidth}
		\includegraphics[width=\textwidth]{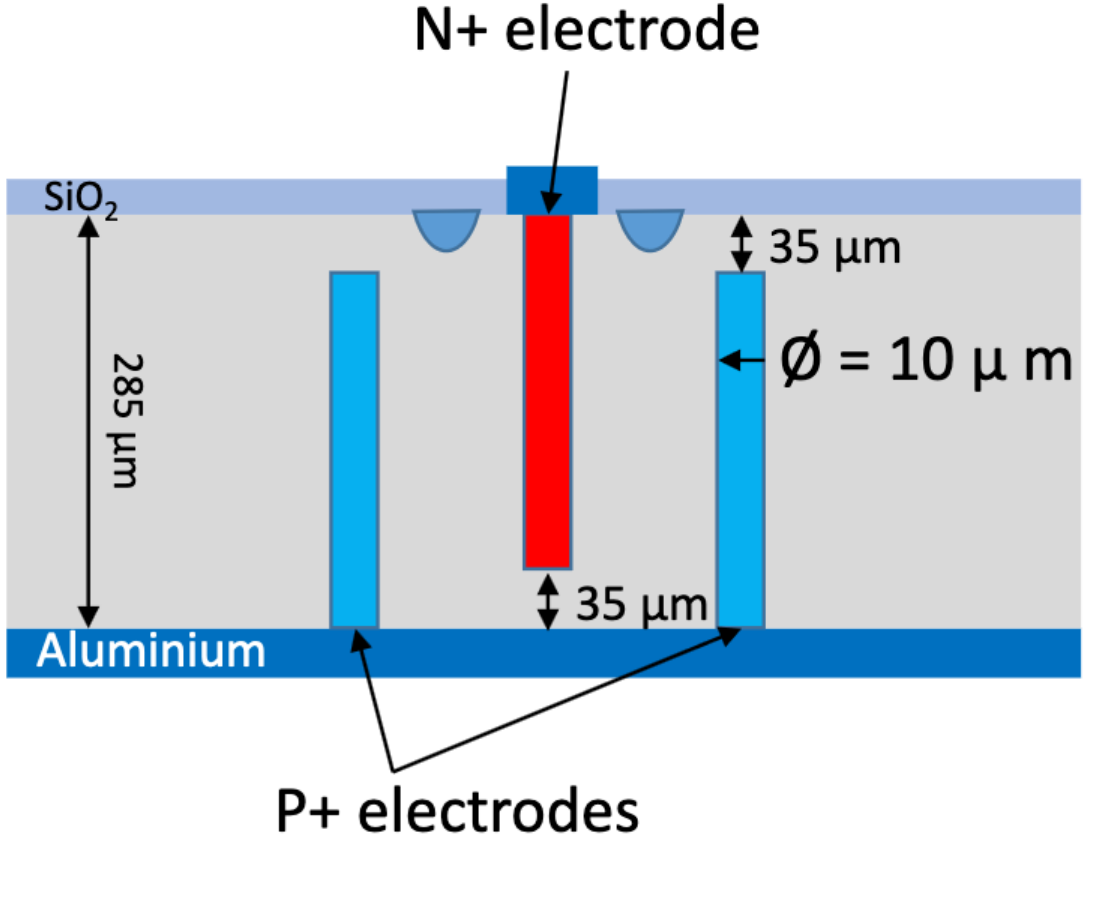}
		\subcaption{}
		\label{fig:3Dscheme}
	\end{subfigure}
	\hspace{0.5cm}
	\begin{subfigure}[f]{0.24\textwidth}
		\centering
		\includegraphics[width=\textwidth]{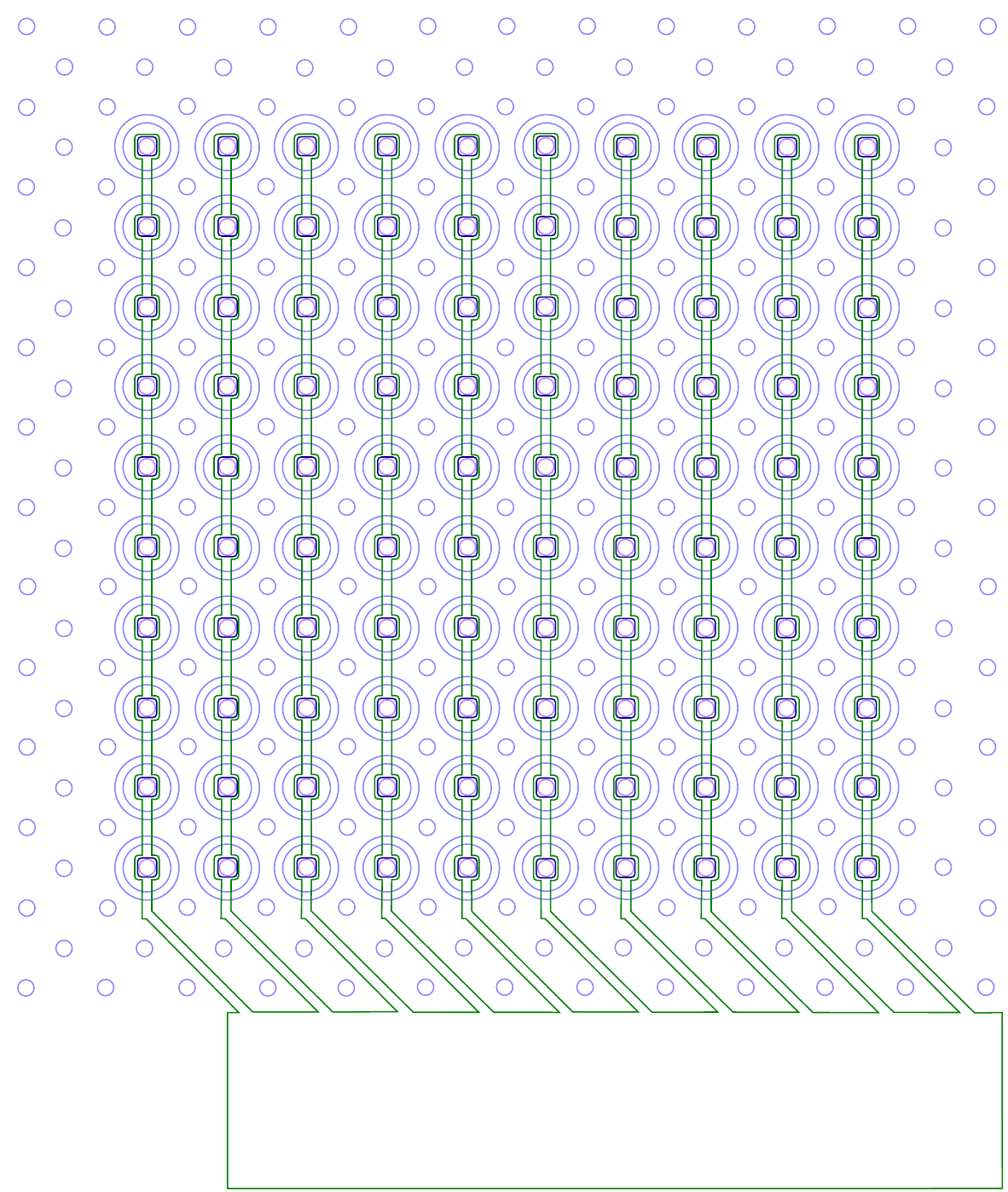}
		\subcaption{}
		\label{fig:6xscheme}
	\end{subfigure}
	\hspace{0.5cm}
	\begin{subfigure}[f]{0.24\textwidth}
		\includegraphics[width=\textwidth]{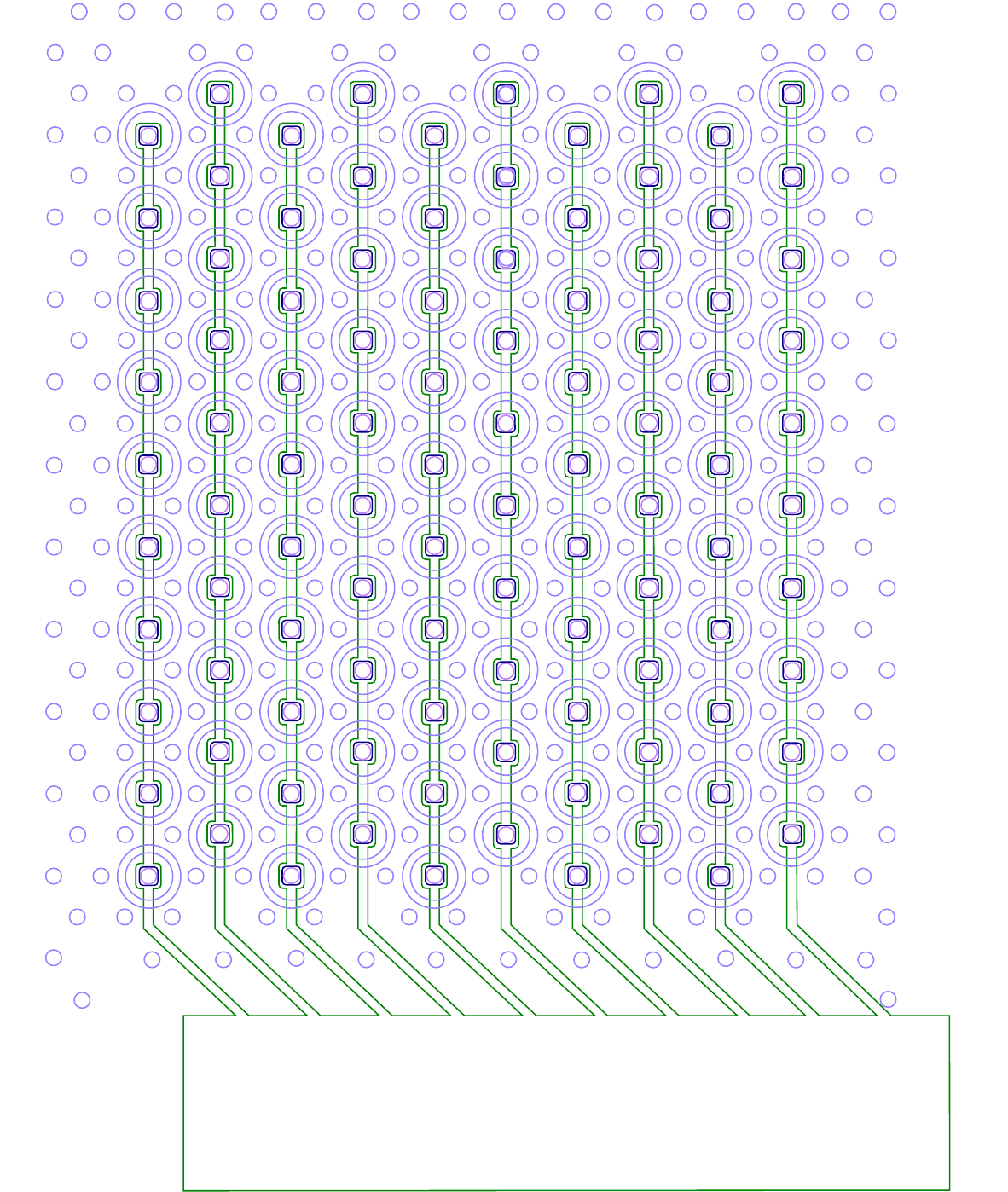}
		\subcaption{}
		\label{fig:7xscheme}
	\end{subfigure}
	\caption{Schematic drawings of the CNM double-sided 3D sensor designs. (a) Cross section of the elemental 3D cell. Layout of the quadratic 6-x (b) and hexagonal 7-x (c) sensors.}
\end{figure}

Dedicated double-sided 3D sensors for timing applications were developed and produced by CNM~\cite{double_sided_3D, Oscar, Terzo} within the common RD50 project RD50-2020-01, "3D detectors optimized for timing applications". The first sensors were delivered in fall 2023. \\ 
A simplified cross section of the sensors is shown in figure~\ref{fig:3Dscheme}. The sensors are \SI{285}{\micro \meter} thick and made of float zone, p-type silicon with a resistivity of more than \SI{5}{\kilo \ohm \centi \meter}. The etched polysilicon columns have a diameter of \SI{10}{\micro \meter} and reach from both sides of the sensor up to a distance of \SI{35}{\micro \meter} to the opposing surface. The ohmic p$^+$ columns from the backside are not directly connected to the backside metallization, but are instead connected through a thin layer of highly doped polysilicon. The n$^+$ readout electrodes on the front are surrounded by circular p-stops. \\
Sensors with two different electrode geometries were fabricated, with either a quadratic (see figure~\ref{fig:6xscheme}) or a hexagonal (see figure~\ref{fig:7xscheme}) layout. The main motivation behind the hexagonal design is an overall reduced distance to the closest readout column and, therefore, improved timing properties \cite{Oscar}. A total of $10\times10$ readout columns are connected to a single bond pad for each sensor. The quadratic sensors are called 6-x, while the hexagonal sensors are named 7-x. The cell size of the quadratic 6-x layout is 50x\SI{50}{\micro \meter}, while the cell diameter is \SI{52}{\micro \meter} for the hexagonal 7-x design. Other, smaller, sensor layouts with fewer readout columns were also produced in both geometries, but are not included in this study.  Single sensors are part of a larger physical structure, called array, consisting of of $4\times4$ sensors. Different arrays are denoted by an additional number after the type name, different sensors by their row and column on the array.\\
Some samples were irradiated with \SI{23}{\mega \electronvolt} protons at the KAZ cyclotron on the KIT Campus North to \SI{1}{\mega \electronvolt} neutron equivalent fluences of \SI{4.11e14}{\per \centi \meter \squared}, \SI{1.06e15}{\per \centi \meter \squared} and \SI{2.9e15}{\per \centi \meter \squared}. For reasons of clarity, the \SI{1}{\mega \electronvolt} neutron equivalent fluence will be denoted as \si{\neqpcm} in the following. The uncertainty on the measured fluences is \SI{15}{\percent} \cite{irradiation}.

\subsection{Experimental setups}
\label{sec:setups}

All IV curves, for unirradiated as well as irradiated sensors, were obtained on a probe station (Karl Suss PA 200) at room temperature ($T=\SI{24}{\celsius}$),  inside a grounded, light-tight, dry air flushed enclosure. The bias voltage and current measurement were provided by a Keithley-237 HV source-measure unit. \\
The top-TCT measurements were carried out on a Particulars large scanning TCT setup at room temperature under a dry atmosphere of less than \SI{10}{\percent} relative humidity. For the measurements, either a red (\SI{640}{\nano \meter}) or an infrared (\SI{1060}{\nano \meter}) Particulars laser was used, with the pulse length set as short as possible (< \SI{0.5}{\nano \second}). The used pulse energy was not calibrated, though a rough estimate arrives at some tens of MIPs of injected charge. The focus of the laser was set to the sensor surface by conducting a z-scan over the metallization connecting the readout columns. The sensors were mounted on a four channel, dual stage amplification readout board designed in Freiburg for fast timing measurements~\cite{TimingBoard}. The first amplification stage is designed after the SCIPP fast timing board and utilizes a SiGe transistor with a bandwidth of \SI{100}{\giga \hertz}. The second stage is an amplifier with \SI{2}{\giga \hertz} bandwidth. Additionally, a temperature sensor for continuous monitoring, and a small hole below the sensor mounting point to reduce material budget for source measurements, are included in the design of the board. The output is recorded by a Tektronix TDS5104, and for each measurement point the area of the waveform averaged over 100 acquisitions is taken.\\
The same fast timing board as described above is also used for the source timing measurements. A collimated Sr-90 source is used to create signal in the sensor. Since this process is randomly distributed in time, a reference detector (REF) is needed in addition to the Device-Under-Test (DUT). To ensure only MIP-like electrons are included in the analysis, a third (trigger) sensor (TRIG) is placed in the setup and a coincident hit in the REF and TRIG sensors is required to record an event. No trigger is set on the DUT, to avoid any biasing of the results, although this results in many empty events due to the geometric constraints of the setup. The reference sensor is a 1.2x\SI{1.2}{\milli \meter \squared} large, \SI{35}{\micro \meter} active thickness HPK LAGD, with a time resolution of $\sigma_{\mathrm{REF}} = \SI{27.0(3)}{\pico \second}$. The TRIG sensor is a larger area FBK LGAD, to increase geometric acceptance and therefore rate of the setup. Both DUT and REF are kept at \SI{-17.0(2)}{\celsius} with PID controlled Peltier elements. All events are recorded with a Tektronix MSO66B with \SI{4}{\giga \hertz} bandwidth and \SI{25}{\giga S} sample rate. The timestamps of the DUT and REF signals are determined using constant fraction discrimination with the threshold set to 50\% of the peak height and the window for the linear fit set to $\pm35\%$ around the threshold. The start of the peak, used to determine all points included in the aforementioned fit window, is defined as the local minimum directly in front of the signal peak. As will be discussed in section~\ref{sec:timing_results}, a SNR cut of 4.5 is applied to the DUT events due to their low SNR, as this results in the lowest relative uncertainty for the time resolution. The difference between the DUT and REF timestamp for each filtered event is binned in a histogram and a gaussian is fitted within a \SI{600}{\pico\second} wide window around the peak of this distribution to obtain its width, the time spread $\sigma_{\mathrm{spread}}$. This window is chosen such that any outlier events with a wrongly converged CFD fit are ignored, while also not cutting into the Gaussian of the time spread distribution. From the time spread, the DUT time resolution can be calculated as $\sigma_{\mathrm{DUT}} = \sqrt{\sigma_{\mathrm{spread}}^2 - \sigma_{\mathrm{REF}}^2}$, with the known $\sigma_{\mathrm{REF}}$.

\section{Results}

\subsection{IV Measurements}
\label{sec:IV}
\begin{figure}[htbp]
	\begin{subfigure}[f]{0.48\textwidth}
		\centering
		\includegraphics[width=\textwidth]{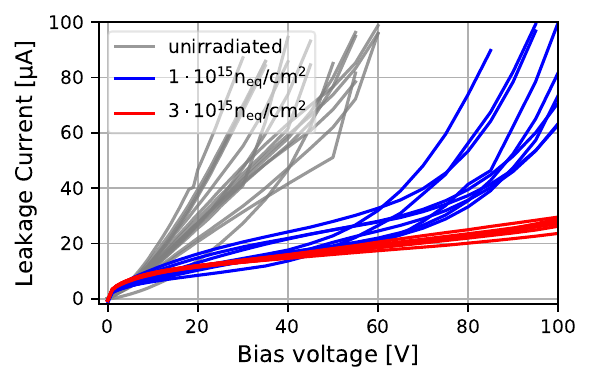}
		\subcaption{}
		\label{fig:IV}
	\end{subfigure}
	\hfill
	\begin{subfigure}[f]{0.48\textwidth}
		\includegraphics[width=\textwidth]{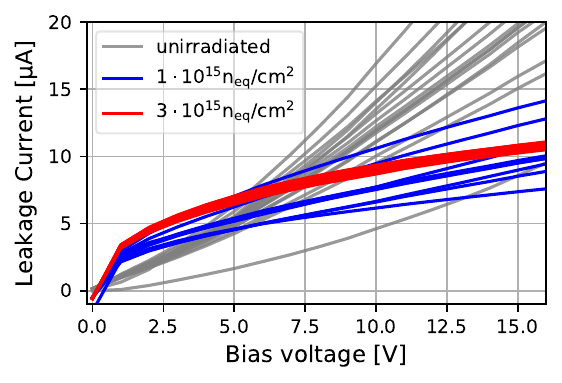}
		\subcaption{}
		\label{fig:IV_close}
	\end{subfigure}
	\caption{IV characteristics of unirradiated as well as irradiated 7-x sensors. The unirradiated and \SI{1e15}{\neqpcm} data stems from the same array before and after irradiation. (a) Entire IV characteristics up to \SI{100}{\volt}. (b) Detail of the IV characteristics at small voltages. Data of unirradiated sensors from~\cite{Iveta}.}
\end{figure}

The basic leakage current behaviour is an essential tool in characterizing the performance of a sensor. In figure~\ref{fig:IV} the IV characteristics of 7-x sensors at different fluence levels is shown. All unirradiated sensors display an overall very high leakage current, especially when considering their volume. Starting at very low voltages, a very steep increase of the current with the voltage can be observed, suggesting a breakdown, as can be seen in \ref{fig:IV_close}. Although all IVs are taken from sensors on the same structure, they show a large variation. The same behaviour can also be observed for all other sensor layouts~\cite{Terzo, Oscar, Iveta}. According to the manufacturer, one wafer from the same production run is not affected by this high leakage current, suggesting a processing issue rather than a problem in the sensor design. Sensors from the unaffected wafer will be investigated in the future. \\
After irradiation, the leakage current decreases strongly, with the exception of low voltages (< \SI{5}{\volt}). There, the leakage current increases faster than in the unirradiated case, but levels out quickly into a more shallow slope. The variation between different sensors decreases, while the breakdown voltage increases with higher fluences. The overall leakage current level is similar for both investigated fluences in the voltage range below \SI{40}{\volt}, before the breakdowns for the \SI{1e15}{\neqpcm} irradiated sensors start. The behaviour after irradiation suggests that the main contribution to the leakage current in the unirradiated sensors does not stem from the sensor bulk and that the defects are healed to a certain degree with the irradiation. The exact origin of the defects and the interaction mechanism with the radiation damages are currently being investigated. 

\subsection{Top-TCT Results}
\label{sec:tct_results}

\begin{figure}[htbp]
	\begin{subfigure}[f]{0.49\textwidth}
		\centering
		\includegraphics[width=\textwidth]{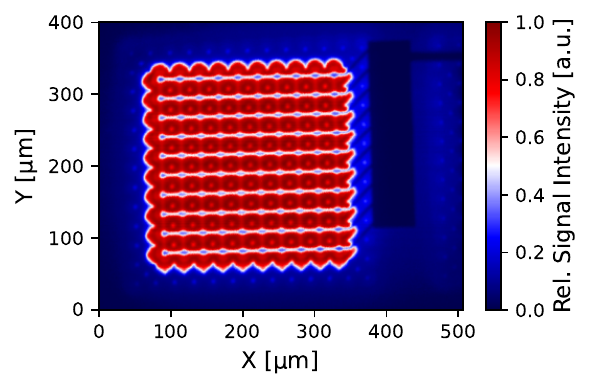}
		\subcaption{}
		\label{fig:6xIR}
	\end{subfigure}
	\hfill
	\begin{subfigure}[f]{0.49\textwidth}
		\includegraphics[width=\textwidth]{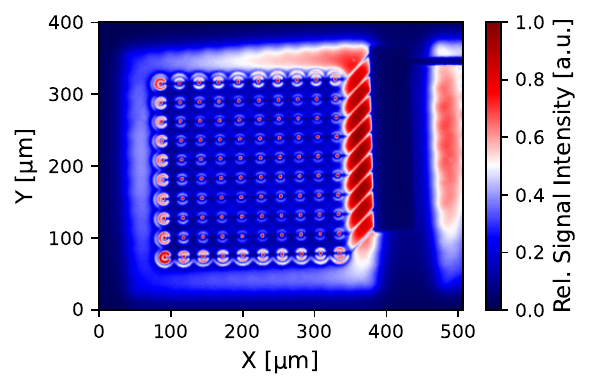}
		\subcaption{}
		\label{fig:6xRed}
	\end{subfigure}
	\caption{Top-TCT scan of an unirradiated 6-x sensor at a bias voltage of \SI{15}{\volt}, including the bondpad region (right side of the image). (a) Scan with infrared laser (\SI{1060}{\nano \meter}). (b) Scan with red laser (\SI{640}{\nano \meter}).}
\end{figure}

In order to investigate the charge collection behaviour over the entire area of the sensor, top-TCT measurements were conducted. By using different wavelength lasers, the behaviour can be studied for charges created in the bulk, using an infrared (\SI{1060}{\nano \meter}) laser, or for charges created directly beneath the surface, utilizing a red (\SI{640}{\nano \meter}) laser. In figure~\ref{fig:6xIR} an infrared scan of a quadratic 6-x sensor biased at \SI{15}{\volt} is shown. The active sensor volume, indicated by a high signal intensity, is separated by horizontal lines of reduced signal, caused by the metallization connecting the readout columns reflecting the laser. Furthermore, the edge of the volume is very irregular, with the outline following the alternating pattern of the p$^+$ ohmic columns. The p$^+$ columns further on the outside are also discernible due to a slightly higher signal intensity than the rest of the inactive volume. A slight substructure is visible in the active sensor volume. At the position of the backside p$^+$ columns, the collected charge is slightly reduced, very likely due to some of the created charges being immediately collected by the electrode, and therefore, not creating a signal through drifting. Additionally, vertical strips of reduced signal between two neighbouring p$^+$ columns can be observed. These inefficiencies are most likely caused by electric field configuration of the investigated sensor type. \\
In figure~\ref{fig:6xRed} the same scan is repeated with a red laser. The most noticeable difference to the infrared scan is that the highest signals are not in the sensor volume, but in the surrounding area occupied by the p$^+$ ohmic columns, especially close to the bondpad region. Interestingly, signal is also created in the area opposite to the sensor. This behaviour hints to most of the charge created close to the surface not reaching the readout electrodes in the bulk, but instead being directly collected by the bond pad and the metal traces, or recombining without creating any signal at all. Furthermore, signal is also created at the holes in the metallization directly above the n$^+$ readout columns, which is not the case for the infrared scan. The circular p-stops surrounding the readout columns are clearly visible, with slightly more charge being collected inside of them. Peculiarly, a significantly larger signal is created in all outermost columns, which is not visible in the infrared scan. This behaviour is also observed in all other layouts, but no satisfactory explanation has been found so far. 

\subsection{Timing Results}
\label{sec:timing_results}

\begin{figure}[htbp]
	\begin{subfigure}[f]{0.49\textwidth}
		\centering
		\includegraphics[width=\textwidth]{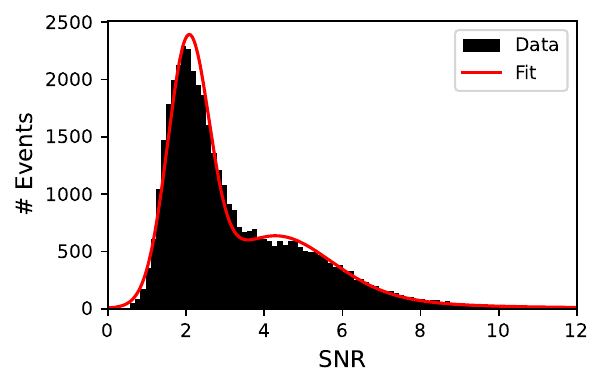}
		\subcaption{}
		\label{fig:SNRfit}
	\end{subfigure}
	\hfill
	\begin{subfigure}[f]{0.49\textwidth}
		\includegraphics[width=\textwidth]{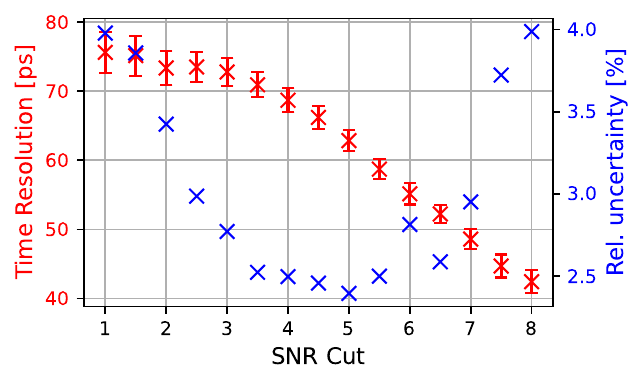}
		\subcaption{}
		\label{fig:resSNR}
	\end{subfigure}
	\caption{Two example plots for the measurement of the \SI{3e15}{\neqpcm} irradiated 7-x-1 R2C3 sensor at a bias voltage of \SI{55}{\volt}. (a) Histogram of the SNR of all events with a time spread within the used fit window. A Gaussian combined with a Langaus distribution, modelling the noise and signal, fitted to the data is shown. (b) Time Resolution and rel. uncertainty of the time resolution plotted against the SNR cut applied to the data.}
\end{figure}

Timing measurements using a radioactive source allow to gauge the timing performance of sensors with MIP-like particles in a laboratory setting without the need for large scale experiments like a test beam. The major drawback of these setups is that, due to scattering of the relatively low energy electrons and the small size of the 3D sensors, many events with no signal in the DUT are recorded. This can be seen in figure~\ref{fig:SNRfit}, where the SNR of all events whose time spread is within the fit window (see section~\ref{sec:setups}) is plotted for one example measurement. The Gaussian peak at low SNR values associated to noise events is clearly visible and significantly larger than the Langaus distribution from real signals. Here, Langaus stands for the convolution of a Landau with a Gaussian distribution. Furthermore, no real separation between noise and signal events is visible, due to the overall low SNR of all events. This low SNR is not exclusive to the hexagonal 7-x layout and is also not caused by irradiation, as it was also already observed in unirradiated sensors. Interestingly, the significantly reduced leakage current for irradiated sensors does not lead to an improved SNR.  \\
In order to not bias the measured time resolution with empty events, a SNR cut is used. Since no clear value can be extracted from figure~\ref{fig:SNRfit} due to the bad separation and no hit efficiency can be determined with this setup, it was instead decided to use the SNR value with the lowest relative uncertainty as the working point. As can be seen in Figure~\ref{fig:resSNR}, the relative uncertainty is roughly parabola shaped and has its minimum around 4.5. The high rel. uncertainty for low SNR can be explained from the fit converging less well due to noise contamination, while for higher SNR the uncertainty increases again due to limited statistics, as most events are cut away. As is expected, the time resolution improves with higher SNR cuts, since higher signal (with the same noise floor) normally correlates with better time resolution.

\begin{figure}[htbp]
	\begin{subfigure}[f]{0.49\textwidth}
		\centering
		\includegraphics[width=\textwidth]{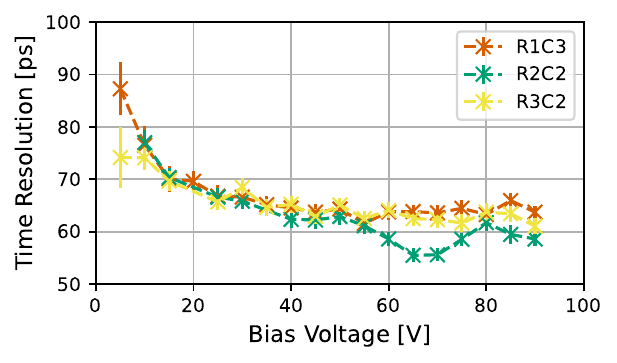}
		\subcaption{}
		\label{fig:1e15res}
	\end{subfigure}
	\hfill
	\begin{subfigure}[f]{0.49\textwidth}
		\includegraphics[width=\textwidth]{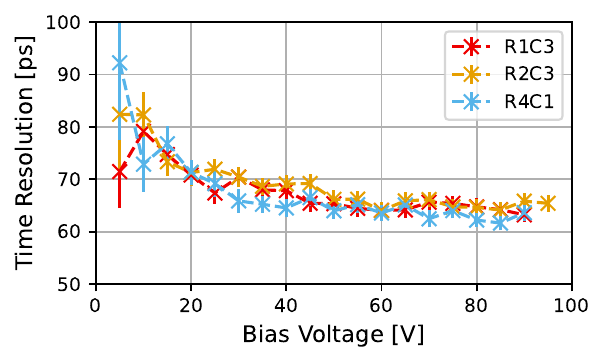}
		\subcaption{}
		\label{fig:3e15res}
	\end{subfigure}
	\caption{Time resolution of different sensors in dependence on the applied bias voltage. (a) Measurements of three sensors on the 7-x-3 array irradiated to \SI{1e15}{\neqpcm}. (b) Measurements of three sensors on the 7-x-1 array irradiated to \SI{3e15}{\neqpcm}}
\end{figure}

The results of time resolution measurements of three different sensors of the 7-x-3 array, irradiated to \SI{1e15}{\neqpcm} are shown in figure~\ref{fig:1e15res}, and in figure~\ref{fig:3e15res} the same is shown for the 7-x-1 array, irradiated to \SI{3e15}{\neqpcm}.
All measured sensors display a similar dependence of the time resolution on the applied bias voltage. With increasing voltage the time resolution improves, as the depleted volume and the electric field strength increases, until it reaches a plateau, which is the case at around \SI{40}{\volt} for the \SI{1e15}{\neqpcm} irradiated sensor and around \SI{55}{\volt} for the \SI{3e15}{\neqpcm} irradiated sensor. For the unirradiated sensor, the plateau is reached slightly earlier at ca. \SIrange{25}{30}{\volt}. \\
The time resolution of the different sensors for each fluence is very similar and agrees within the measurement uncertainty for most voltages, with the only exception being the R2C2 sensor on the 7-x-1 array at voltages above \SI{60}{\volt}, where a decrease in time resolution can be observed. This decrease coincides with a decrease in noise, and also jitter, while the mean SNR stays constant. In the same voltage range the break down occurs (see figure \ref{fig:IV}). A change in time resolution at the break down voltage was also observed in unirradiated sensors, but always lead to an increase instead of a decrease. Further investigation of this behaviour is currently ongoing. The overall time resolution reached by sensors of both fluences at voltages above \SI{55}{\volt} is very similar. The time resolution measured for unirradiated 7-x sensors with a slightly older readout board configuration and a SNR cut of 5 was \SI{61.6(9)}{\pico \second}~\cite{Iveta}, and therefore comparable to the resolution measured for the irradiated sensors. In publications about the performance of previous designs of 3D sensors, time resolutions of down to \SI{25}{\pico \second} after irradiation were measured~\cite{Diehl}. The better values reported there can be explained by the investigated structures being single columns, instead of 10x10 column arrays, which have a significantly lower capacitance and therefore also lower expected time resolution. Smaller layouts of the sensors investigated in this work have also displayed better time resolutions~\cite{Iveta}. Investigations of the same double-sided 3D sensor layouts with a TPA-TCT setup~\cite{comparison_3D} resulted in time-of-arrival differences between \SIrange{20}{55}{\pico\second} for different positions inside one unit cell, which is smaller than the time resolution reported here.

\section{Conclusion and Outlook}
\label{sec:outlook}

3D sensors are one of the main technologies with the ability to deliver the time resolution needed for future collider experiments without compromising on other important characteristics like radiation hardness or spatial resolution. In this study, a first characterisation of (quadratic) 6-x and (hexagonal) 7-x unirradiated as well as irradiated double-sided 3D sensor structures produced by CNM was conducted, utilizing IV, top-TCT and source timing measurements. The overall leakage current of the sensors is high, but decreases strongly after irradiataion, while the break down voltage increases. The reason for this behaviour is currently still under investigation. Infrared top-TCT scans show a mostly homogenous charge collection in the active volume of the sensor, although some finer substructure due to the ohmic columns from the backside and the electric field configuration is visible. Surface charge injection reveals an interesting pattern of higher collected charge around the bondpad and the outermost columns. This effect is currently being investigated further. Although the investigated sensors display a low SNR and an unsatisfactory separation between noise and signal, their overall timing resolution of ca. \SI{65}{\pico \second} after irradiation is promising, since it is only marginally worse than for the unirradiated sensor and the structures are large, therefore having a higher capacity, compared to previously investigated 3D sensors with a better time resolution. \\
Further (timing-)TCT and source timing measurements of other (irradiated) layouts are currently being conducted to compare their performance and find other outliers like the 7-x-3 R2C2 sensor (see figure~\ref{fig:1e15res}). Test beam measurements are currently being prepared, with the main focus being the hit detection efficiency and the spatially resolved measurements of timing parameters.

\acknowledgments

This work was partially funded by BMFTR under grant 05H24VF.\\
The proton irradiations were carried out at the KAZ cyclotron operated by the company ZAG located on the KIT Campus North.


 \bibliographystyle{JHEP}
 \bibliography{biblio.bib}

\end{document}